\documentclass[showkeys,aps,pra,10pt,showpacs,amsmath,amssymb,floatfix,superscriptaddress,longbibliography,twocolumn]{revtex4-2}

\usepackage{physics}
\usepackage{graphicx}
\usepackage[usenames,dvipsnames,svgnames,table]{xcolor}
\usepackage{tcolorbox}
\usepackage[unicode=true,
            pdfusetitle,
            bookmarks=true,
            bookmarksnumbered=false,
            bookmarksopen=false,
            breaklinks=true,
            pdfborder={0 0 0},
            backref=false,
            colorlinks=true,
            hypertexnames=false]{hyperref}
\hypersetup{linkcolor=NavyBlue,urlcolor=NavyBlue,citecolor=NavyBlue}
\usepackage{bbm}

\begin{document}

\title{Joint optimal measurement for locating two incoherent optical point sources near the Rayleigh distance}

\author{Yingying Shi}
\author{Xiao-Ming Lu}
\email{lxm@hdu.edu.cn}
\affiliation{School of Sciences, Hangzhou Dianzi University, Hangzhou 310018, China}

\begin{abstract}
Simultaneously optimizing the estimation of the centroid and separation of two incoherent optical point sources is constrained by a tradeoff relation determined by an incompatibility coefficient.  
At the Rayleigh distance the incompatibility coefficient vanishes and thus the tradeoff relation no longer restricts the simultaneous optimization of measurement for a joint estimation.
We construct such a joint optimal measurement by an elaborated analysis on the operator algebra of the symmetric logarithmic derivative.
Our work not only confirms the existence of a joint optimal measurement for this specific imaging model, but also gives a promising method to characterize the condition on measurement compatibility for general multiparameter estimation problems.  
\end{abstract}


\maketitle

\section{Introduction}
The resolution power of an imaging system is an important problem in optics~\cite{Dekker1997,Villiers2016}.
The conventional approach on measuring the resolution power is the Rayleigh criterion, which gives a minimum distance that two incoherent optical point sources can be distinguished in vision~\cite{LordRayleigh1879}.
A modern approach on the resolution power can be established in terms of parameter estimation theory, where the resolution power is measured by the estimation precision of the separation between two optical point sources~\cite{Ram2006,Chao2016}.
By using quantum parameter estimation theory~\cite{Helstrom1976,Paris2009,Zhang2022}, which takes into consideration the optimization over quantum measurements, Tsang \textit{et al}. showed that the resolution power for two incoherent optical point sources in the sub-Rayleigh region can be grossly improved by the spatial mode demultiplexing (SPADE) measurement~\cite{Tsang2016b,Tsang2019a}.

The SPADE measurement is optimal for estimating the separation of two point sources but needs the prior information about the centroid of the two point source to align the involved spatial modes~\cite{Tsang2016b}.
On the other hand, direct imaging is good at estimating the centroid of two point sources but performs poorly for estimating the separation.     
The tradeoff between the measurement efficiencies for estimating the centroid and the separation is a manifestation of Heisenberg's uncertainty principle and can be analyzed with resorting to the information regret tradeoff relation (IRTR)~\cite{Lu2021}.

The IRTR restricts the simultaneous optimization for the centroid estimation and the separation estimation through an incompatibility coefficient~\cite{Shao2022}.
For a spatially-invariant imaging system with a Gaussian point-spread function, this incompatibility coefficient approaches to its maximum value as the separations goes to zero and vanishes for very large separations.
Besides, the incompatibility coefficient has a nontrivial zero point when the separation is double the standard deviation of the intensity distribution, which we called the Rayleigh distance.
This opens up the possibility of existing a measurement that is simultaneously optimal for estimating the centroid and the separation.

In this work, we will give a positive answer to the existence of a joint optimal measurement for estimating the centroid and the separation at the nontrivial zero point of the incompatibility coefficient.
For quantum multiparameter estimation, a complete characterization of the condition on the existence of a measurement that is optimal for estimating all parameter remains open, e.g., see Ref.~\cite{Yang2019b,Chen2022a,Chen2022}.
However, it is known that the eigenvectors of the symmetric logarithmic derivative (SLD) operator with respect to a parameter constitutes the basis of an optimal measurement for estimating that parameter~\cite{Braunstein1994}.
The optimal measurement constructed in this way in general depends on the true value of the parameters to be estimated and thus cannot be directly applied when possible values of the parameters lie in a wide range~\cite{Braunstein1996}.
Nevertheless, such an optimal measurement is still useful, as it reveals the fundamental limit of estimation precision and can be approached with adaptive feedback control~\cite{Fujiwara2006,Hayashi2005}.
We will show that for the parameter estimation model of locating two incoherent optical point sources, the SLD operators with respect to the centroid and to the separation are both not unique.
Although the ``standard'' SLD operators considered in the previous work~\cite{Tsang2016b} are not commuting with each other, we can find another pair of commuting SLD operators at the nontrivial zero point of the incompatibility coefficient.
Furthermore, we will construct a joint optimal measurement in terms of the common eigenvectors of the commuting SLD operators.

We organized this paper as follows.
In Sec.~\ref{sec:estimation}, we give a brief introduction on quantum multiparameter estimation theory and the measurement incompatibility problem. 
In Sec.~\ref{sec:superresulotion}, we describe the necessary background about the joint estimation problem for locating two incoherent optical point sources.
In Sec.~\ref{sec:joint_optimal}, we investigate the joint optimal measurement for estimating both the centroid and the separation when the incompatibility coefficient vanishes.
We summarize our work in Sec.~\ref{sec:conclusion}.

\section{Quantum multiparameter estimation theory} 
\label{sec:estimation}

Let us start with a brief introduction on quantum multiparameter estimation and measurement incompatibility. 
Assume that the state of a quantum system depends on a set \(\theta = (\theta_1, \theta_2,\ldots,\theta_n)\) of \(n\) unknown parameters and is described by a parametric family \(\rho_\theta\) of  density operator. 
The values of \(\theta\) is estimated through processing the outcomes of a measurement performed on the quantum system.
A quantum measurement can be mathematically described by a positive operator valued measure (POVM) \(M = \qty{E_x \mid E_x \ge 0, \sum_x E_x = \mathbbm{1} }\), where \(\mathbbm 1\) denotes the identity operator and \(x\) the measurement outcomes.
The probability of obtaining a measurement outcome \(x\) is  \(p(x) = \tr(\rho _\theta E_x)\) according to Born's rule in quantum mechanics.
The data processing is represented by the estimators  \(\hat \theta = (\hat\theta_1,\hat\theta_2, \ldots, \hat\theta_n)\), which are maps from the observation data to the estimates. 
The error-covariance matrix of any unbiased estimator obeys the Cram\'er-Rao bound~\cite{Kay1993,Wasserman2010}
\begin{equation}
    \mathrm{Cov}(\hat\theta) \ge N^{-1} F(M)^{-1},
\end{equation} 
where \(N\) is the number of experimental repetitions and \(F(M)\) is the Fisher information matrix (FIM) defined by
\begin{equation}
    [F(M)]_{j,k} = \sum_x p(x)^{-1} \pdv{p(x)}{\theta_j} \pdv{p(x)}{\theta_k}. 
\end{equation} 
The FIM depends on the quantum measurement, meaning that we can optimize over quantum measurements.
In quantum estimation theory~\cite{Helstrom1969,Helstrom1967,Helstrom1968,Braunstein1994,Liu2020}, it is known that the FIM is bounded from above by the quantum FIM \(\mathcal F\) in the sense that \(F(M) \leq \mathcal F\), i.e., \(\mathcal F - F(M)\) is positive semi-definite, for all POVMs \(M\).
The quantum FIM is defined by
\begin{equation}
    {{\cal F}_{j,k}} = \mathrm{Re}\,\tr( {{\rho _\theta }{L_j}{L_k}}),
\end{equation}
where the SLD operators ${L_j}$ for ${\theta _j}$ are Hermitian operators satisfying 
\begin{equation} \label{eq:SLD}
    \pdv{\rho_\theta}{\theta_j}= \frac{1}{2} (L_j \rho_\theta + \rho_\theta L_j).
\end{equation}
Note that the SLD operator might be not uniquely determined by the above equality;
This fact will play an important role in our work.

In general, the entries of the FIM cannot be simultaneous maximized by optimizing over quantum measurements;
This is known as the measurement compatibility problem in quantum multiparameter estimation theory~\cite{Ragy2016,Carollo2019a,Suzuki2019,Chen2022,Chen2022a}.
The degree of nonoptimality of a measurement for estimating an individual parameter \(\theta_j\) can be measured by the square-rooted and normalized version of \emph{information regret}~\cite{Lu2021}: 
\begin{equation}
    \Delta_j = \qty( \frac{\mathcal F_{j,j} - [F(M)]_{j,j}}{\mathcal F_{j,j} })^{1/2}.
\end{equation} 
As a manifestation of Heisenberg's uncertainty principle, the information regrets for any two different parameters, e.g., \(\theta_j\) and \(\theta_k\), must obey the IRTR~\cite{Lu2021}:
\begin{equation} \label{eq:IRTR}
    \Delta_j^2 + \Delta_k^2 + 2 \sqrt{1 - c_{j,k}^2} \Delta_j \Delta_k \ge c_{j,k}^2,
\end{equation} 
where \(c_{j,k}\) is the incompatibility coefficient defined as 
\begin{equation} \label{eq:incompatibility}
    c_{j,k} = \frac{ \tr| \sqrt\rho [L_j, L_k] \sqrt\rho |}{ 2 \sqrt{\mathcal F_{j,j} \mathcal F_{k,k}} }
\end{equation}
with \(|X|:=\sqrt{X^\dagger X}\) for any operator \(X\).
The IRTR is tight for when \(\rho\) is pure, meaning that there always exist a quantum measurement such that the equality in Eq.~\eqref{eq:IRTR} holds.
For mixed states, it remains open whether the IRTR is tight, even for the special case where the incompatibility coefficient \(c_{j,k}\) vanishes.

\section{Estimation problem for locating two incoherent point sources}
\label{sec:superresulotion}

\begin{figure}[bpt]
  \begin{center}
  \includegraphics[scale=0.4]{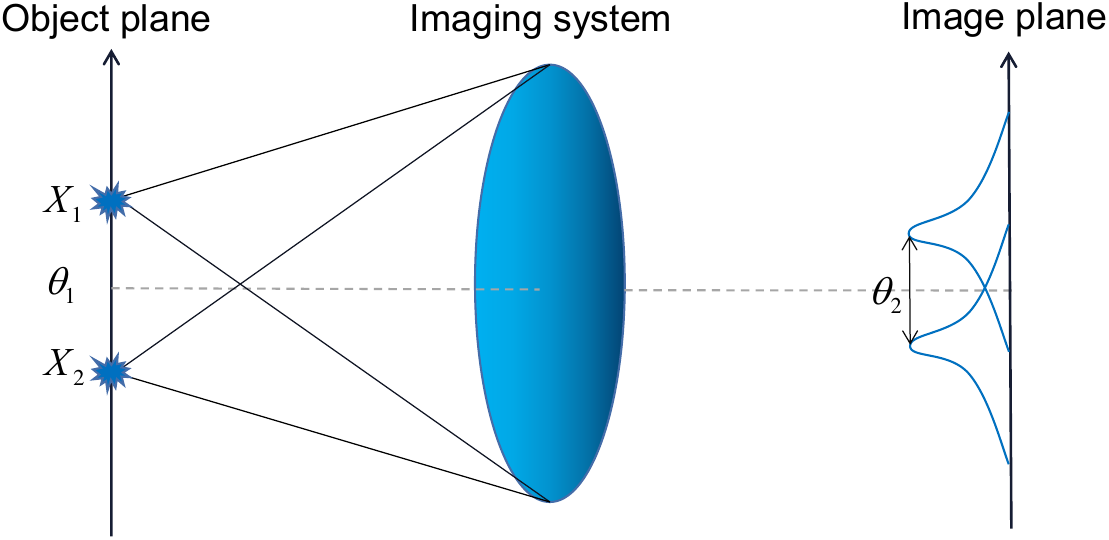}
  \caption{ \label{fig:model}
  Model of resolving two incoherent optical point sources. 
  Here, $X_1$ and $X_2$ are the 1-dimensional coordinates of the first and second point sources, respectively.
  The parameters $\theta_1$ and \(\theta_2\) represent the centroid and the separation of the two point sources, respectively.}
  \end{center}
\end{figure}

We now consider a concrete multiparameter estimation problem---the joint estimation of the centroid and the separation of two incoherent optical point sources~\cite{Tsang2016b}.
The model is illustrated in Fig.~\ref{fig:model}.
Following Ref.~\cite{Tsang2016b}, the quantum states of the spatial degree of a photon arriving at the imaging can be described by the density operator 
\begin{equation}
    \rho = \frac{1}{2} \qty(\op{\psi_1} + \op{\psi_2}),
\end{equation} 
where \(\ket{\psi_j}\) denotes the quantum state of a photon from the \(j\)th source.
Denoted by \(\psi(x)\) the normalized point-spread function of the imaging sources.
The state vector \(\ket{\psi_j}\) can be expressed in the coordinate basis as 
\begin{equation}
	\ket{\psi_j} = \int \dd x\, \psi(x-X_k) \ket x,
\end{equation}
where \(\ket x\) is the photon image-plane position eigenket.
The parameters to be estimated the centroid \(\theta_1\) and the separation \(\theta_2\) that are defined as
\begin{equation}
	\theta _1 = ( X_1 + X_2) / 2 
	\qand 
	\theta _2 = X_2 - X_1.
\end{equation}

The SLD equation Eq.~\eqref{eq:SLD} can be solved by resorting to represent all relevant operator by matrix with an orthonormal basis.
A specific orthonormal basis for the minimal subspace that supports both \(\rho\)  and its derivatives with respect to \(\theta_1\) and \(\theta_2\) is given by~\cite{Tsang2016b}
\begin{align}
	\ket{e_1} & = \frac{1}{\sqrt{2 (1 - \delta)}} \qty(\ket{\psi_1} - \ket{\psi_2}), \nonumber\\
	\ket{e_2} & = \frac{1}{\sqrt{2 (1 + \delta)}} \qty(\ket{\psi_1} + \ket{\psi_2}), \nonumber\\
	\ket{e_3} & = \frac{1}{\eta_3}\qty[
  		\frac{1}{\sqrt2}\qty(\pdv{\ket{\psi_1}}{X_1} + \pdv{\ket{\psi_2}}{X_2}) - \frac{\gamma}{\sqrt{1-\delta}} \ket{e_1}
  	], \nonumber\\
  	\ket{e_4} & = \frac{1}{\eta_4}\qty[
  		\frac{1}{\sqrt2}\qty(\pdv{\ket{\psi_1}}{X_1} - \pdv{\ket{\psi_2}}{X_2}) + \frac{\gamma}{\sqrt{1+\delta}} \ket{e_2}
  	], \label{eq:e_basis}
\end{align}
where the coefficients \(\delta\), \(\kappa\), \(\gamma\), and \(\beta\) are defined as 
\begin{align}
  	\delta 	& = \int \dd x\,\psi(x-X_1) \psi(x-X_2), \nonumber\\
	\kappa  & = \int_{-\infty}^\infty \dd x\, \qty[ \pdv{\psi(x)}{x}]^2, \nonumber\\
	\gamma 	& = \int_{-\infty}^\infty \dd x\, \pdv{\psi(x)}{x} \psi(x - \theta_2), \nonumber\\ 
	\beta 	& = \int_{-\infty}^\infty \dd x\, \pdv{\psi(x)}{x} \pdv{\psi(x - \theta_2)}{x},
\end{align}
and \(\eta_3\) and \(\eta_4\) are determined by the normalization condition as
\begin{align}
	\eta_3 	= \sqrt{\kappa + \beta - \frac{\gamma^2}{1 - \delta}}, \quad
	\eta_4 	= \sqrt{\kappa - \beta - \frac{\gamma^2}{1 + \delta}}.
\end{align}
With this orthonormal basis, the density operator for the image-plane one-photon state is represented by
\begin{equation} \label{eq:rho}
    \rho  = \frac{1 - \delta }{2} \op{e_1} + \frac{1 + \delta}{2}  \op{e_2}.
\end{equation} 
The SLD operators with respect to  \(\theta_1\) and  \(\theta_2\) are represented by
\begin{align}
	L_1 & = \mqty(
		0 & \frac{2\gamma\delta}{\sqrt{1-\delta^{2}}} & 0 & \frac{2\eta_4}{\sqrt{1-\delta}}\\
		\frac{2\gamma\delta}{\sqrt{1-\delta^{2}}} & 0 & \frac{2\eta_3}{\sqrt{1+\delta}} & 0\\
		0 & \frac{2\eta_3}{\sqrt{1+\delta}} & 0 & 0\\
		\frac{2\eta_4}{\sqrt{1-\delta}} & 0 & 0 & 0
	), \nonumber\\
	L_2 & = \mqty(
		\frac{-\gamma}{1-\delta} & 0 & \frac{-\eta_3}{\sqrt{1-\delta}} & 0\\
		0 & \frac{\gamma}{1+\delta} & 0 & \frac{-\eta_4}{\sqrt{1+\delta}}\\
		\frac{-\eta_3}{\sqrt{1-\delta}} & 0 & 0 & 0\\
		0 & \frac{-\eta_4}{\sqrt{1+\delta}} & 0 & 0
	). \label{eq:sld12}
\end{align}
With the above matrices, the incompatibility coefficient defined by Eq.~\eqref{eq:incompatibility} is given by~\cite{Shao2022}:
\begin{equation} \label{eq:c}
	c^2 = \frac{\beta^2}{\kappa(\kappa - \gamma^2)}.
\end{equation}
Here, we omit the subscripts of \(c\) for brevity, as we here only consider the estimation of two parameters.

\begin{figure}
    \centering
    \includegraphics{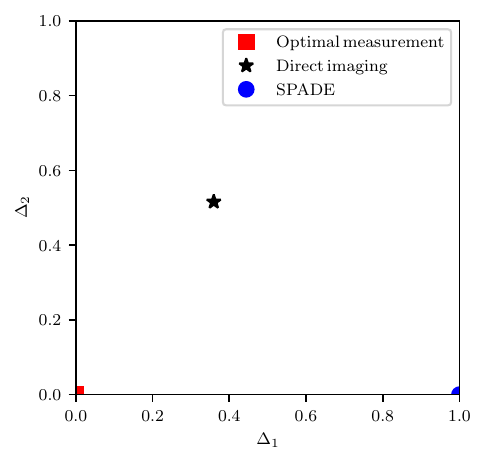}
    \caption{\label{fig:regrets}
        Information regrets for the centroid \(\theta_1\) and the separation \(\theta_2\) of direct imaging, the SPADE measurement, and the ideal optimal joint measurement.
        Here, the true value of the separation is set to \(\theta_2=2\sigma\).
        The SPADE is performed with Hermite-Gaussian modes, whose origin point is placed at the centroid of the two sources. 
    }
\end{figure}

We are interested in the case where the incompatibility coefficient \(c\) becomes zero; This will happen when \(\beta\) vanishes according to Eq.~\eqref{eq:c}.
Assume that the point-spread function of the imaging system is Gaussian, viz., 
\begin{equation}
    \psi(x) = (2\pi\sigma^2)^{- 1 / 4} \exp(-\frac{x^2}{4\sigma^2}),
\end{equation} 
where $\sigma$ is a characteristic width on the image plane. 
In such a case, it can be shown that~\cite{Shao2022}
\begin{equation}
  \beta = \frac{4\sigma^2 - \theta_2^2}{16\sigma^2} \exp(- \frac{\theta_2^2}{8\sigma^2}).  
\end{equation} 
The nontrivial case of \(\beta=0\) occurs at \(\theta_2 = 2\sigma\), which we call the Rayleigh distance.
At this parameter point, the IRTR becomes \(\Delta_1^2 + \Delta_2^2 \geq 0\), which no longer restrict the simultaneous optimization of quantum measurements for estimating the centroid and separation of two incoherent optical point sources.
This gives us the possibility of an optimal joint measurement.
If there exists such a measurement, it will outperform direct imaging and the SPADE measurement in the perspective of multiparameter estimation (see Fig.~\ref{fig:regrets} for an illustration).

\section{Joint optimal measurement}
\label{sec:joint_optimal}

We now derive the joint optimal measurement for the aforementioned two-parameter estimation problem.
Braunstein and Caves~\cite{Braunstein1994} showed that, for single parameter estimation, the POVM constituted by the eigen-projectors of the SLD operator is optimal in the sense that the extracted classical Fisher information attains its quantum limit---the quantum Fisher information.
For multiparameter estimation, if the SLD operator commutes with each other, they have a common eigen-projectors, which constitutes a joint optimal measurement in the sense that the classical Fisher information matrix under this measurement equals to the quantum Fisher information matrix.
We will utilize this property to construct a joint optimal measurement for the problem of jointly estimating the centroid and separation of two incoherent optical point sources.

Although the SLD operators given in Eq.~\eqref{eq:sld12} do not commute with each other, it is possible to find a pair of commuting SLD operators, as the SLD operators might not be uniquely determined by an estimation problem.
It is evident that, if \(L_j\) is an SLD operator of \(\rho\) with respect to \(\theta_j\) and \(K_j\) is a Hermitian operator satisfying \(K_j \rho = \rho K_j = 0\), then \(L_j'=L_j+ K_j\) is also a Hermitian operator satisfying the SLD equation Eq.~\eqref{eq:SLD} and thus can be considered as an SLD operator.
The additional \(K_j\) to the SLD operator does not affect the value of the QFI matrix as \(\tr( L_j \rho L_k) = \tr[(L_j + K_j)\rho (L_k + K_k)]\).

To find a set \(\qty{K_j}\) of Hermitian operators such that the SLD operators commutes with each other, we will analyze in detail the representations of the SLD operator.
Let \(\mathcal H\) be the (local) relevant Hilbert subspace of the estimation model at a parameter point \(\theta\).
Concretely, \(\mathcal H\) is the union of the support of the density operators \(\rho\) and its derivatives \(\partial\rho/\partial \theta_j\) at the parameter point \(\theta\).
Note that the relevant Hilbert subspace \(\mathcal H\) depends on the value of \(\theta\).
We only need to consider the SLD operators acting on \(\mathcal H\). 
Furthermore, the Hilbert subspace can be decomposed as \(\mathcal H = \mathcal S \oplus \mathcal K\), where \(\mathcal S\) denotes the support of the density operator \(\rho\) and \(\mathcal K\) is the orthogonal complement of \(\mathcal S\) in \(\mathcal H\).
We decompose the matrix representation of the SLD operators on \(\mathcal H\) in the block form with respect to the space decomposition \(\mathcal H = \mathcal S \oplus \mathcal K\):
\begin{align} \label{eq:sld_block_form}
	L_j = \mqty(
		A_j & B_j \\
		B_j^\dagger & K_j
	).
\end{align}
Because \(L_j\) is Hermitian operators, the matrices \(A_j\) and \(K_j\) are all Hermitian.
The matrices \(A_j\) and \(B_j\) are uniquely determined by the SLD equation~\eqref{eq:SLD}, whereas \(K_j\) can be an arbitrary Hermitian matrix.

For the two-parameter estimation problem, we need to find the Hermitian matrices \(K_1\) and \(K_2\) such that \(L_1 L_2 = L_2 L_1\).
Using the above block forms, we get
\begin{equation}
	L_1 L_2 = \mqty(
		A_1 A_2 + B_1 B_2^\dagger& 
		A_1 B_2 + B_1 K_2 \\
		B_1^\dagger A_2 + K_1 B_2^\dagger
		& B_1^\dagger B_2 + K_1 K_2
	).
\end{equation}
Comparing block by block both sides of the condition \(L_1 L_2 = L_2 L_1\), we obtain the following equations:
\begin{align}
	A_1 A_2 - A_2 A_1 	&= B_2 B_1^\dagger - B_1 B_2^\dagger, \label{eq:requirement0} \\
  	B_1 K_2 - B_2 K_1	&= A_2 B_1 - A_1 B_2, \label{eq:requirement1} \\
  	K_1 K_2 - K_2 K_1 &= B_2^\dagger B_1 - B_1^\dagger B_2. \label{eq:requirement2} 
\end{align}
Here, Eq.~\eqref{eq:requirement0} is a necessary condition on the existence of a joint optimal measurement.
Equations~\eqref{eq:requirement1} and \eqref{eq:requirement2} represent the conditions that should by satisfied by \(K_j\).

For the concrete estimation problem considered in this paper, the set \(\qty{\ket{e_j} \mid j=1,2,3,4}\) given in Eq.~\eqref{eq:e_basis} is an orthonormal basis of the relevant Hilbert subspace \(\mathcal H\).
Meanwhile, the subspace \(\mathcal S\) and \(\mathcal K\) are spanned by \(\qty{\ket{e_1}, \ket{e_2}}\) and \(\qty{\ket{e_3}, \ket{e_4}}\), respectively.
With this  Hilbert space decomposition, we can extract the matrices \(A_1\), \(B_1\), \(A_2\), and \(B_2\) from the SLD operators given in Eq.~\eqref{eq:sld12}.
It can be verified that the necessary condition Eq.~\eqref{eq:requirement0} is satisfied.
For the canonical SLD operator in Eq.~\eqref{eq:sld12}, both \(K_1\) and \(K_2\) are the zero matrix but can be set to arbitrary Hermitian matrices.
Our objective is to find a pair of \(K_1\) and \(K_2\) that satisfy the conditions Eqs.~\eqref{eq:requirement1} and \eqref{eq:requirement2}.

We can expand an arbitrary 2-dimensional matrix \(X\) as the linear combination of the 2-dimensional identity matrix \(\hat\sigma_0\) and the Pauli matrices \(\hat\sigma_1\), \(\hat\sigma_2\), and \(\hat\sigma_3\), namely, 
\begin{equation}
	X = \sum_{\alpha=0}^3 v_\alpha(X) \hat\sigma_\alpha
	\qq{with}
	v_\alpha(X) = \frac12 \tr(\hat\sigma_\alpha X).
\end{equation}
If \(X\) is Hermitian, \(v_\alpha(X)\) are all real numbers.
Note that both sides of Eq.~\eqref{eq:requirement2} is traceless so that the expansion coefficients \(v_0\) of bosh sides vanish.
Therefore, the conditions Eq.~\eqref{eq:requirement1} and Eq.~\eqref{eq:requirement2} give a system of \(7\) equations involving \(8\) variables, viz., \(v_\alpha(K_1)\) and \(v_\alpha(K_2)\) for \(\alpha=0,1,2,3\).
This system of equations is undetermined and has infinitely many solutions.
We give a specific solution as follows:
\begin{align}
    K_1 &= \qty(\frac{2 \gamma}{1-\delta^2} - \frac{2\kappa}{\gamma}) \hat\sigma_0 
    - \frac{2\delta\gamma}{1-\delta^2} \hat\sigma_3, \\
    K_2 &= \frac{\eta_3\eta_4}{\gamma} \hat\sigma_1 + \frac{(1+\delta^2)\gamma}{1-\delta^2} \hat\sigma_3. 
\end{align}

\begin{figure}[tb]
	\centering
	\includegraphics[]{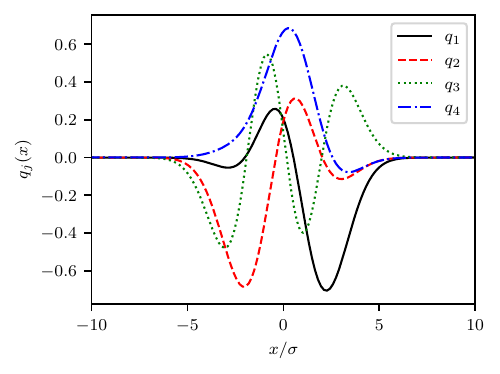}
	\caption{The wave functions of the measurement basis for the joint optimal measurement given in Eq.~\eqref{eq:joint_optimal}.}
	\label{fig:optimal_wavefun}
\end{figure}

Putting the above \(K_1\) and \(K_2\) into Eq.~\eqref{eq:sld_block_form}, we get a pair of commuting SLD operators \(L_1\) and \(L_2\), which can be simultaneously diagonalized.
Denote by \(\phi_j\) the \(j\)th common eigenvectors of \(L_1\) and \(L_2\) and \(\phi_{j,k}\) the \(k\)th component of \(\phi_j\).
The measurement basis for the joint optimal measurement is given by 
\begin{equation} \label{eq:joint_optimal}
	\ket{q_j} \equiv \sum_{k=1}^4 \phi_{j,k} \ket{e_k} \qfor j=1,2,3,4.
\end{equation}
With Eq.~\eqref{eq:e_basis}, we can express the wave function of the measurement basis \(\ket{q_j}\) in terms of the point-spread function and its derivative.
We numerically solve the common eigenvectors of \(L_1\) and \(L_2\) at the Rayleigh distance \(\theta_2 = 2\sigma\) and plot in Fig.~\ref{fig:optimal_wavefun} the wave function \(q_j(x) = \ip{x}{q_j}\) of the joint optimal measurement basis.

We here emphasize that the optimal measurement basis constructed by the eigenvectors of the  SLD operator in general depends on the true values of the parameters to be estimated.
The significance of such a measurement basis is that it can reveal the fundamental limit of the joint estimation precision with a quantum measurement.
For practical application, the adaptive feedback control can help the quantum measurement approach to the optimal status.

\section{Conclusions} \label{sec:conclusion}

In this work, we have constructed a joint optimal measurement that simultaneously extracts the maximal Fisher information about the centroid and the separation of two incoherent optical point sources.
The method we used is utilizing the fact that the SLD operator is not uniquely determined by the parameter estimation model.
By decomposing the SLD operators into the block form, we have found a pair of commuting SLD operators whose common eigenvectors can be taken as the basis of the joint optimal measurement.
Our work, on the one hand, confirms the existence of a joint optimal measurement for the specific model of simultaneously estimating the centroid and the separation of two incoherent optical point sources at the Rayleigh distance, on the other hand, gives a promising method to characterize the condition on measurement compatibility for general multiparameter estimation problems.  

\begin{acknowledgments}
This work is supported by the National Natural Science Foundation of China (Grants No. 12275062, No. 11935012, and No. 61871162).
\end{acknowledgments}

\end{document}